\documentclass[aps,prd,superscriptaddress,notitlepage,showkeys,showpacs,10pt]{revtex4-1}

\usepackage{graphicx}
\usepackage{amsmath}
\usepackage{amssymb}
\usepackage{color}
\usepackage{bm}
\usepackage{hyperref}
\usepackage{float}

\definecolor{purple}{rgb}{0.8,0,0.6}

\begin{document}
\title{Anomaly-driven inverse cascade and inhomogeneities in a magnetized chiral plasma in  the early Universe}

\author{E. V. Gorbar}
\affiliation{Department of Physics, Taras Shevchenko National Kiev University, Kiev, 03022, Ukraine}
\affiliation{Bogolyubov Institute for Theoretical Physics, Kiev, 03680, Ukraine}

\author{I. Rudenok}
\affiliation{Department of Physics, Taras Shevchenko National Kiev University, Kiev, 03022, Ukraine}

\author{I. A. Shovkovy}
\affiliation{College of Integrative Sciences and Arts, Arizona State University, Mesa, Arizona 85212, USA}
\affiliation{Department of Physics, Arizona State University, Tempe, Arizona 85287, USA}

\author{S. Vilchinskii}
\affiliation{Department of Physics, Taras Shevchenko National Kiev University, Kiev, 03022, Ukraine}
\affiliation{Universit\'e de Gen\'eve, D\'epartement de Physique Th\'eorique, CH-1211 Gen\'eve 4, Switzerland}

\begin{abstract}
By making use of a simple model that captures the key features of the anomalous Maxwell equations, 
we study the role of inhomogeneities on the evolution of magnetic fields in a chiral plasma. We find that 
inhomogeneities of the chiral asymmetry by themselves do not prevent the anomaly-driven inverse cascade and, 
as in the homogeneous case, the magnetic helicity is transferred from shorter to longer wavelength helical modes 
of the magnetic field. However, we also find that the evolution appears to be sensitive to the effects of diffusion. 
In the case when diffusion is negligible, the inverse cascade slows down considerably compared to the homogeneous 
scenario. In the case of the primordial plasma, though, we find that the diffusion is substantial and efficiently suppresses 
chiral asymmetry inhomogeneities. As a result, the inverse cascade proceeds practically in the same way as 
in the chirally homogeneous model. 
\end{abstract}



\maketitle

\section{Introduction}

It is well known that nonzero magnetic fields are present in all parts of the observed Universe, including
galaxies, clusters of galaxies, and the intergalactic medium. However, the ultimate origin of such 
magnetic fields remains largely unknown \cite{Kronberg,Rubinstein,Giovannini-review,Vallee,Durrer}. 
The observed magnetic fields in galaxies and clusters of galaxies, which are of the order of a microgauss 
in strength, can be explained by an amplification of pre-existing weak seed magnetic fields via well 
established astrophysical processes (e.g., the dynamo and/or the flux-conserving compression during the 
gravitational collapse) accompanying the structure formation. As for the seed magnetic fields, they 
must be generated by a different mechanism. In general, there could exist two principal classes of 
models for the origin of the seed magnetic fields: (i) those connected with the processes before
the structure formation in the early Universe, and (ii) those connected with the processes that accompany 
the gravitational collapse associated with the structure formation.

The recent observation of the magnetic fields of the order of $B \sim 10^{-15}~\mbox{G}$ in the voids 
\cite{Neronov,Ando,Tavecchio,Dolag}, which are unlikely to be generated by any known astrophysical 
mechanisms, strongly supports the idea of the primordial origin of the magnetic fields \cite{Kandus,Widrow} and, 
perhaps, identifies the cosmological magnetic fields as the required seeds for the generation of the observed 
galactic magnetic fields. Various scenarios of magnetogenesis during the inflation, reheating, electroweak and 
strong phase transitions were proposed and studied in the literature \cite{Rubinstein,Giovannini-review,Durrer,Kandus,Widrow}.
The main problem of magnetogenesis in the early Universe is that the corresponding magnetic fields become 
quickly diluted during the cosmic expansion. As a result, they are unlikely to survive and serve as 
seeds for the generation of the galactic magnetic fields. This reasoning is based on the standard treatment of
the evolution of magnetic fields in magnetohydrodynamics of a finite-resistivity relativistic plasma
\cite{Jedamzik:96,Subramanian:97,Banerjee:03,Banerjee:04,Giovannini:2003yn,Kahniashvili:2010gp}.
While short wavelength modes are quickly erased by the magnetic diffusion due to the finite electric 
conductivity of the plasma, viscosity plays the dominant role in dissipation at larger scales \cite{Banerjee:03,Banerjee:04}.

In addition, it is well known that magnetic fields produced after the inflation (e.g., generated at the electroweak 
epoch) quickly become short-ranged (deeply sub-horizon) due to the cosmological 
expansion and the field dissipation~\cite{Durrer:2003ja,Caprini:2009pr,Saveliev:12}. 
However, for helical magnetic fields, in view of the turbulence effects, a part of their energy can be transferred from 
shorter to longer wavelengths (``inverse cascade") and, thus, escape the dissipation due to viscous and 
resistive effects \cite{Christensson:2002xu,Banerjee:04,Durrer:2003ja,Campanelli:07}.

As recently demonstrated in Ref.~\cite{Boyarsky,Tashiro:2012mf,Manuel:2015zpa,Hirono:2015rla}, 
there is another mechanism of the inverse cascade in a magnetized chiral plasma which may profoundly 
change the evolution of magnetic fields in the early Universe. The argument is based on the idea that 
an instability countering the dissipation can 
develop as a result of an Abelian anomaly in the action \cite{Vilenkin:80a,Joyce:97,Giovannini:1997eg,
Shaposhnikov,Giovannini} that would lead to the generation of the hypercharge magnetic fields in the 
primordial plasma, which in turn gives rise to ordinary magnetic fields via the electroweak mixing. 
This process was believed to be limited to temperatures $T\gtrsim 80$~TeV, as the chirality flipping reactions at 
lower temperatures mix left- and right-handed particles and, thus, quickly erase the chiral asymmetry 
in the plasma. In Ref.~\cite{Boyarsky} it was shown, however, that a non-linear interaction 
between the chiral asymmetry and helical magnetic field may prolong the evolution significantly, i.e.,  
until the temperature drops down to about a few MeV, despite the presence of the chirality flipping reactions and
the magnetic diffusion.  Moreover, the same interaction induces an effective mechanism for transferring energy from 
the magnetic modes at dissipation-affected short scales to long-lived modes with long wavelengths. The 
underlying phenomenon is similar to the inverse cascade in ordinary magnetohydrodynamics 
\cite{Christensson:2002xu}, but driven by anomalous transport properties in the chiral plasma, rather than the 
turbulence. The corresponding anomalous transport properties were first obtained in the framework of the strongly 
coupled $N=4$ super-Yang-Mills theory using the AdS/CFT correspondence \cite{Banerjee:2008th,Loganayagam:2008is,
Rangamani:2009xk} and, later, derived in general relativistic theories with anomalous chiral and/or conformal symmetries
\cite{Son:2009tf,Kharzeev:2011vv,Kalaydzhyan:11,Dubovsky:2011sj,Amado:2011zx,Neiman:10}. 

The mechanism proposed in Ref.~\cite{Boyarsky} links the evolution of the magnetic field helicity with 
the divergence of the fermion axial current and the chiral magnetic effect \cite{Fukushima:08,Kharzeev}. 
The latter implies, in particular, that an imbalance between the number density of the right- and 
left-handed fermions (described semirigorously by the chiral chemical potential $\mu_5$) in a 
magnetic field $\mathbf{B}$ leads to an induced electric current, $\mathbf{j}=e\mu_5\mathbf{B}/(2\pi^2c)$. 
Although magnetic field modes with nonzero wave vectors were considered, only the space-averaged 
chiral anomaly equation and a constant chiral chemical potential were utilized in the analysis of Ref.~\cite{Boyarsky}
that revealed the inverse cascade.
Later, by analyzing the anomalous Maxwell equations, the dynamical evolution of the magnetic 
fields and the chiral fermion imbalance were studied in more detail in Ref.~\cite{Tashiro:2012mf,Manuel:2015zpa,Hirono:2015rla}. Depending 
on the chosen initial conditions, it was found that the helicity can be transferred  either from the fermions 
to the magnetic fields or vice versa.

By taking into account that the helical electromagnetic fields are necessarily inhomogeneous, it is natural 
to expect that the chiral asymmetry should also develop inhomogeneities during the inverse cascade.
Therefore, the anomalous Maxwell equations of Ref.~\cite{Boyarsky} should be modified to allow for such a 
dependence. Within hydrodynamics, the spatial dependence of the chiral chemical potential was included  
in Ref.~\cite{Boyarsky:2015faa}, where it was shown that the inverse cascade survives also in the spatially 
inhomogeneous case. The same generic conclusion is reached is a study of real-time dynamics of chirally 
imbalanced Dirac fermions using a classical statistical field theory approach \cite{Buividovich:2015jfa}.

In order to study the effects of an inhomogeneous chiral chemical potential self-consistently, one can 
also try to use the framework of the chiral kinetic theory. By utilizing an expansion in powers of electromagnetic 
fields and derivatives, a closed set of anomalous Maxwell equations and the expressions for the electric 
and axial currents were derived for an inhomogeneous chiral plasma \cite{inhomogeneous}. Additionally,
the equations for the electric charge chemical potential and chiral chemical potential were obtained. The corresponding 
results of  Ref.~\cite{inhomogeneous} motivated the current study that addresses the fate of the inverse cascade in the 
chirally asymmetric inhomogeneous primordial plasma. For the sake of simplicity, we neglect the chirality 
flipping processes by arguing as in Ref.~\cite{Boyarsky} that both the chiral asymmetry and the magnetic helicity 
should survive in the primordial plasma on the time-scales much longer than the diffusion or the chirality 
flipping times (till temperatures $10-100~\mbox{MeV}$).

This paper is organized as follows. We review the anomalous Maxwell equations for a homogeneous relativistic 
plasma in Sec.~\ref{CKT}. Their generalization for a chirally asymmetric inhomogeneous relativistic plasma are 
considered in Sec.~\ref{inhomogeneous-plasma}. Our numerical results are presented in Sec.~\ref{numerical-results}.
The summary and conclusions are given in Sec.~\ref{Conclusion}.

\section{Anomalous Maxwell equations for chirally homogeneous plasma}
\label{CKT}

Before considering an inhomogeneous plasma, let us start by introducing the 
notation and recalling the key details of the inverse cascade in the case of a homogeneous 
chiral plasma \cite{Boyarsky,Tashiro:2012mf,Manuel:2015zpa,Hirono:2015rla}. The electromagnetic fields in the plasma satisfy the usual
Maxwell's equations
\begin{eqnarray}
\mathbf{\nabla}\times\textbf{B}&=& 4\pi e\textbf{j}+ \frac{\partial\textbf{E}}{\partial t}, \quad\quad
\mathbf{\nabla}\times\textbf{E}=-\frac{\partial\textbf{B}}{\partial t},
\label{Maxwell_Eqs}\\
\mathbf{\nabla}\cdot\textbf{E}&=&4\pi en, \quad\quad
\mathbf{\nabla}\cdot\textbf{B}=0,
\label{Maxwell_Eqs_2}
\end{eqnarray}
where the electric current is given by
\begin{equation}
\label{j0}
e\textbf{j}=\frac{\alpha }{2\pi^2}\mu_5\textbf{B}+\sigma\textbf{E}
\end{equation}
and the charge density $en$ vanishes in view of the overall neutrality of the plasma.
In the expression for the current, the first term describes the usual Ohm's current, while 
the second term (proportional to the chiral chemical potential $\mu_5$) captures the signature 
property of the chirally asymmetric plasma, i.e., the chiral magnetic effect \cite{Fukushima:08,Kharzeev}. 
By substituting the expression for the current into the Maxwell's equations, we derive the 
following equation for the magnetic field:
\begin{equation}
\label{eq_B_x}
\frac{\partial\textbf{B}}{\partial t}=\frac{1}{4\pi \sigma}
\left(\mathbf{\nabla}^2\textbf{B}-\frac{\partial^2\textbf{B}}{\partial t^2}
+\frac{2\alpha }{\pi} \mu_5 [\mathbf{\nabla}\times\textbf{B}]\right).
\end{equation}
Here $\sigma$ is the electrical conductivity and $\alpha=  e^2 $ (we use units with $\hbar=1$ and $c=1$) 
is the fine-structure constant. The evolution of the spatially homogeneous chiral chemical potential is determined 
from the chiral anomaly relation averaged over the space volume $V$, i.e.,
\begin{equation}
\label{cae}
\frac{\partial\mu_5}{\partial t}=\frac{3\alpha }{2\pi^2 T^2}\frac{1}{V}\int d^3x \textbf{E}\cdot\textbf{B}
=-\frac{3\alpha}{4\pi^2 T^2}\frac{d\mathcal{H}}{dt},
\end{equation}
where $T$ is temperature,
\begin{equation}
\mathcal{H}(t)=\frac{1}{V}\int d^3x\textbf{A}\cdot\textbf{B}
\end{equation}
is the magnetic helicity, and $\textbf{A}$ is the vector potential. For later use, it is useful to express the magnetic helicity as an integral 
in momentum space
\begin{equation}
\mathcal{H}=\frac{1}{V}\int\frac{d^3k}{(2\pi)^3}\textbf{A}_{\textbf{k}}\cdot\textbf{B}_{\textbf{k}}^*\equiv\int dk\mathcal{H}_k,
\end{equation}
where $\textbf{A}_{\textbf{k}}$ and $\textbf{B}_{\textbf{k}}$ are the Fourier transforms of the vector potential and magnetic field, 
respectively.

In the context of cosmology, it is convenient to rewrite Eqs.~\eqref{eq_B_x} and \eqref{cae} using the conformal coordinates 
suitable for the description of the primordial relativistic plasma in the Friedmann-Robertson-Walker (FRW) expanding 
Universe. The corresponding dimensionless conformal time is defined \cite{Boyarsky} as $\eta={M^*}/{T}$, 
where $M^*=\sqrt{\frac{90}{8\pi^3g^*}}M_{Pl}$, $M_{Pl}$ is the Planck mass, $g^*$ is the effective number of the
relativistic degrees of freedom, and $T$ is the temperature of the Universe (although we will use in the equations below the
conformal time, we denote it $t$). Here we tacitly assume the
relation between the scale factor $a$ and temperature as in the radiation-dominated era: $a(t)={1}/{T}$. The 
Maxwell equations in the FRW Universe acquire \cite{Giovannini-review,Banerjee:04,Joyce:97} their 
standard flat-space form after the following rescaling: $\mu_5\rightarrow a\mu_5$, $\sigma\rightarrow a\sigma$, 
$T\rightarrow aT$, $\textbf{B}\rightarrow a^2\textbf{B}$, and $\textbf{E}\rightarrow a^2\textbf{E}$. Note that 
the dimensionless electrical conductivity equals $\sigma\approx 70$ in the primordial plasma \cite{Baym}.

As follows from Eqs.~(\ref{eq_B_x}) and (\ref{cae}), the Fourier components of the magnetic helicity 
satisfy the following set of equations: 
\begin{eqnarray}
\label{scalar-H}
\frac{\partial\mathcal{H}_k}{\partial t} &=&-\frac{k^2}{2\pi\sigma}\mathcal{H}_k
+\frac{\alpha}{\pi^2}\frac{k\mu_5}{\sigma}\mathcal{H}_k, 
\\
\label{scalar-mu}
\frac{\partial\mu_5}{\partial t}&=&-\frac{3\alpha}{4\pi^2} \int dk\frac{\partial\mathcal{H}_k}{\partial t}.
\end{eqnarray}
These equations were studied in Ref.~\cite{Boyarsky}, where it was shown that an inverse cascade 
takes place transferring the energy from the magnetic field modes with large $k$ to the modes with small $k$. 
As explained in Introduction, this mechanism may provide the seeds of magnetic fields 
on cosmologically large scales.

In the next section, we will see how the equations for the evolution of magnetic field and chiral asymmetry 
are modified in the presence of space coordinates dependent chiral chemical potential.

\section{Chirally inhomogeneous plasma}
\label{inhomogeneous-plasma}

A closed set of the Maxwell's equations for an inhomogeneous chiral plasma was derived in Ref.~\cite{inhomogeneous},
using the chiral kinetic theory in the relaxation time approximation. In essence, they are the same as in 
Eq.~(\ref{Maxwell_Eqs}), but the expressions for the current and charge densities contain a number of new 
dissipative and nondissipative (topological) terms. To the second order in derivatives, the corresponding 
expressions in the high-temperature limit (i.e., $|\mu_{\lambda}|\ll T$, where 
$\mu_\lambda=\mu+\lambda\mu_5$ and $\lambda=\pm 1$ for the fermions of the left and right chirality, respectively) 
take the form \cite{inhomogeneous}:
\begin{eqnarray}
\label{j}
\textbf{j} &\simeq & \frac{e\textbf{B}\mu_5}{2\pi^2}+\frac{\tau T^2}{9}\left(e\textbf{E}-\frac{\partial\mu}{\partial \textbf{x}}\right)
-\frac{e\tau^2T^2}{9}\frac{\partial \textbf{E}}{\partial t},
\\
\label{jn}
n &\simeq & \frac{T^2\mu}{3}-\frac{\tau T^2}{3}\frac{\partial\mu}{\partial t}+\frac{\tau^2T^2}{9}(\mathbf{\nabla}^2\mu
-e\mathbf{\nabla}\cdot\textbf{E})-\frac{e\tau}{2\pi^2}\left(\textbf{B}\cdot\frac{\partial\mu_5}{\partial\textbf{x}}\right),
\end{eqnarray}
where $\tau$ is the relaxation time parameter. 
The electric charge chemical potential $\mu$ and chiral chemical potential $\mu_5$ are governed by the equations
\begin{eqnarray}
\label{mu}
\frac{\partial\mu}{\partial t}+\frac{3}{2\pi^2 T^2}e\textbf{B}\cdot\frac{\partial\mu_5}{\partial\textbf{x}}
-\frac{\tau}{3}\left(\mathbf{\nabla}^2\mu-e\mathbf{\nabla}\cdot\textbf{E}\right)=0,\\
\label{mu_5}
\frac{\partial\mu_5}{\partial t}+\frac{3}{2\pi^2 T^2}e\textbf{B}\cdot\frac{\partial\mu}{\partial\textbf{x}}
-\frac{\tau}{3}\mathbf{\nabla}^2\mu_5=\frac{3e^2\textbf{E}\cdot\textbf{B}}{2\pi^2T^2}.
\end{eqnarray}
As we will see in our numerical calculations and is natural to expect, the electric fields are much smaller 
than the magnetic fields during the evolution of chirally asymmetric plasma.
Therefore, it is justified to neglect all spacetime derivatives of the electric field in Eqs.~(\ref{j})-(\ref{mu}). 
Further, Eq.~(\ref{mu}) implies that the time derivative of the 
electric chemical potential is of the second order in electromagnetic field and derivatives. Additionally,
in view of the electric neutrality of the plasma, we choose the electric chemical potential to be vanishing initially.
By taking into account that the corresponding chemical potential enters Eqs.~(\ref{j}) and (\ref{mu_5}) via the
gradient, the corresponding terms in the equations are of the third order in the electromagnetic field and 
derivatives and, consequently, can be omitted. Then, for the electric current density, we have the following 
approximate expression:
\begin{equation}
\label{j_fin}
e\textbf{j}\simeq\frac{\alpha}{2\pi^2}\mu_5\textbf{B}+\sigma\textbf{E},
\end{equation}
where we expressed the relaxation time parameter through the electrical conductivity $\tau=9\sigma/(e^2T^2)$. 
By substituting current (\ref{j_fin}) into the Maxwell's equation Eq.~(\ref{Maxwell_Eqs}),
we obtain the following equation for the magnetic field:
\begin{equation}
\label{eq_B_x_new}
\frac{\partial\textbf{B}}{\partial t}=\frac{1}{4\pi\sigma}\left(\mathbf{\nabla}^2\textbf{B}
+\frac{2\alpha }{\pi}[\mathbf{\nabla}\times\mu_5\textbf{B}]\right).
\end{equation}
Obviously, this equation is consistent with Eq.~(\ref{eq_B_x}) obtained in the case of a homogeneous chiral plasma. (Note that 
the second order time derivative of the magnetic field can be neglected in a highly conducting plasma.)
In the same approximation, the chiral chemical potential satisfies the following equation:
\begin{equation}
\frac{\partial\mu_5}{\partial t}-\frac{3\sigma}{\alpha}\mathbf{\nabla}^2\mu_5=\frac{3\alpha\textbf{E}\cdot\textbf{B}}{2\pi^2 T^2},
\label{evolution-equation-mu-5}
\end{equation}
This latter differs from the corresponding equation (\ref{cae}) for the homogeneous chiral plasma in two ways. 
First of all, the right-hand side of Eq.~(\ref{cae}) is the space-averaged expression on the right-hand side of
Eq.~(\ref{evolution-equation-mu-5}). Secondly, Eq.~(\ref{evolution-equation-mu-5}) contains the additional diffusion 
term on its left-hand side.

In the case of a homogeneous chiral plasma, the description of the inverse cascade was most natural 
in the momentum space representation \cite{Boyarsky}. In fact, the simplest realization of the cascade was already possible in 
a model with a relatively small number of helical magnetic modes characterized by a discrete set of wave vectors. 
Below, we use a similar approach also for the description of an inhomogeneous chiral plasma. Just like in the 
homogeneous model, it is convenient to rewrite Eqs.~(\ref{eq_B_x_new}) and (\ref{evolution-equation-mu-5}) in 
the momentum space and use an approximation with a finite (relatively small) number of modes with discrete 
wave vectors.

We study the system in a finite box of dimensions $L\times L\times L$. The discrete coordinates and wave vectors 
are given by 
\begin{eqnarray}
x=\frac{m_x L}{N},\quad y=\frac{m_y L}{N},\quad z=\frac{m_z L}{N},\quad m_x,m_y,m_z=0,1,...,N-1 ,\\
k_x=\frac{2\pi n_x}{L},\quad k_y=\frac{2\pi n_y}{L},\quad k_z=\frac{2\pi n_z}{L},\quad n_x,n_y,n_z=0,1,...,N-1,
\end{eqnarray}
respectively. The Fourier and inverse Fourier transforms for magnetic field and chiral chemical potential are given by
\begin{eqnarray}
\textbf{B}(\textbf{x})&=&\frac{1}{V}\sum\limits_\textbf{k}e^{i\textbf{k}\textbf{x}}\textbf{B}_\textbf{k},\qquad\qquad
\textbf{B}_\textbf{k}=\frac{V}{N^3}\sum\limits_\textbf{x}e^{-i\textbf{k}\textbf{x}}\textbf{B}(\textbf{x}),
\\
\label{fourier_mu_5_x}
\mu_5(\textbf{x})&=&\frac{1}{V}\sum\limits_\textbf{k}e^{i\textbf{k}\textbf{x}}\mu_{5,\textbf{k}},\qquad\qquad
\mu_{5,\textbf{k}}=\frac{V}{N^3}\sum\limits_\textbf{x}e^{-i\textbf{k}\textbf{x}}\mu_5(\textbf{x}),
\end{eqnarray}
where $\textbf{x}=\{x,y,z\}$ and $\textbf{k}=\{k_x,k_y,k_z\}$.

By making use of the discrete momentum space representation defined above, we derive the following 
dimensionless form of the equations for the magnetic field and chiral chemical potential:
\begin{eqnarray}
\label{B_k_disc_main}
\frac{\partial\textbf{B}_\textbf{k}}{\partial t} &=& \frac{1}{4\pi\sigma}\left(-k^2\textbf{B}_\textbf{k}
+\frac{2i\alpha }{\pi}\frac{1}{V}\sum\limits_{\textbf{k}_1}\mu_{5,\textbf{k}_1}[\textbf{k}\times\textbf{B}_{\textbf{k}-\textbf{k}_1}]
\right),
\\
\label{mu_5_k_disc_main}
\frac{\partial\mu_{5,\textbf{k}}}{\partial t}  &=&  \frac{3\alpha}{8\pi^3\sigma V}\left(\sum\limits_{\textbf{k}_1}i[\textbf{k}_1
\times\textbf{B}_{\textbf{k}_1}]\textbf{B}_{\textbf{k}-\textbf{k}_1}-\frac{2\alpha }{\pi}\frac{1}{V}\sum\limits_{\textbf{k}_1}
\sum\limits_{\textbf{k}_2}\mu_{5,\textbf{k}_1}\textbf{B}_{\textbf{k}_2}\textbf{B}_{\textbf{k}-\textbf{k}_1-\textbf{k}_2}\right)
-\frac{3\sigma}{\alpha}\textbf{k}^2\mu_{5,\textbf{k}},
\end{eqnarray}
where $V$ is a volume of the system. Note that we used the following 
expression for the electric field:
\begin{equation}
\textbf{E} =\frac{1}{4\pi \sigma} \left(\mathbf{\nabla}\times \textbf{B}-\frac{2\alpha }{\pi} \mu_5 \textbf{B}\right),
\end{equation}
which results from Eqs.~(\ref{Maxwell_Eqs}) and (\ref{j0}) with the time derivative of the electric field 
neglected. Equations~(\ref{B_k_disc_main}) and (\ref{mu_5_k_disc_main}) is a closed system that describes a self-consistent interplay of the magnetic 
field modes and inhomogeneous chiral asymmetry in the primordial plasma. Note that the homogeneous 
case is reproduced when only the $\textbf{k}=0$ mode of the chiral chemical potential, $\mu_{5,0}$, is 
retained in the analysis. In the next section, we will study these equations numerically.

\section{Numerical results}
\label{numerical-results}

In order to solve Eqs.~\eqref{B_k_disc_main}--\eqref{mu_5_k_disc_main} numerically, we consider a simplified 
model with $k_x=k_y=0$ and a discrete set of wave vectors $k_z$ (with $N=10$ nodes). The magnetic helicity 
is defined as $\mathcal{H}_k=\frac{1}{V}\frac{k}{2\pi^2}(|B_k^+|^2-|B_k^-|^2)$, where $B_k^+=B_{k,x}-iB_{k,y}$ and 
$B_k^-=B_{k,x}+iB_{k,y}$. Following the approach of Refs.~\cite{Boyarsky,Tashiro:2012mf,Manuel:2015zpa,Hirono:2015rla}, the initial values of the 
magnetic field are chosen to satisfy the condition of the maximal helicity $|B_k^-|=0$.  Also, as in Ref.~\cite{Boyarsky}, 
we assume that the modes with larger wave vectors $|\textbf{k}|$ initially carry larger fractions of the magnetic 
helicity $H_{k}$. 

\begin{figure}[t!]
\includegraphics[width=0.47\textwidth]{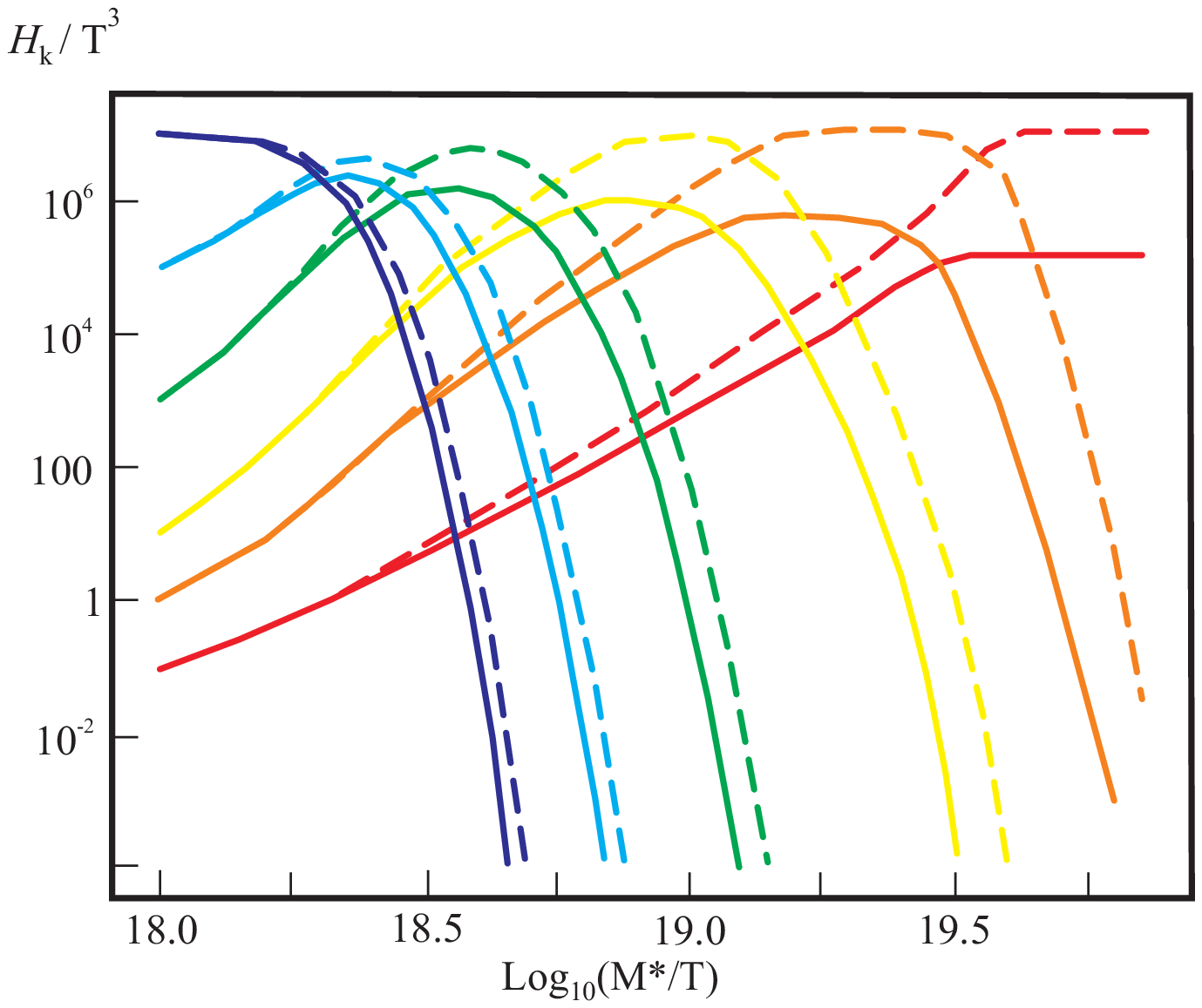}\hspace{0.05\textwidth}%
\includegraphics[width=0.48\textwidth]{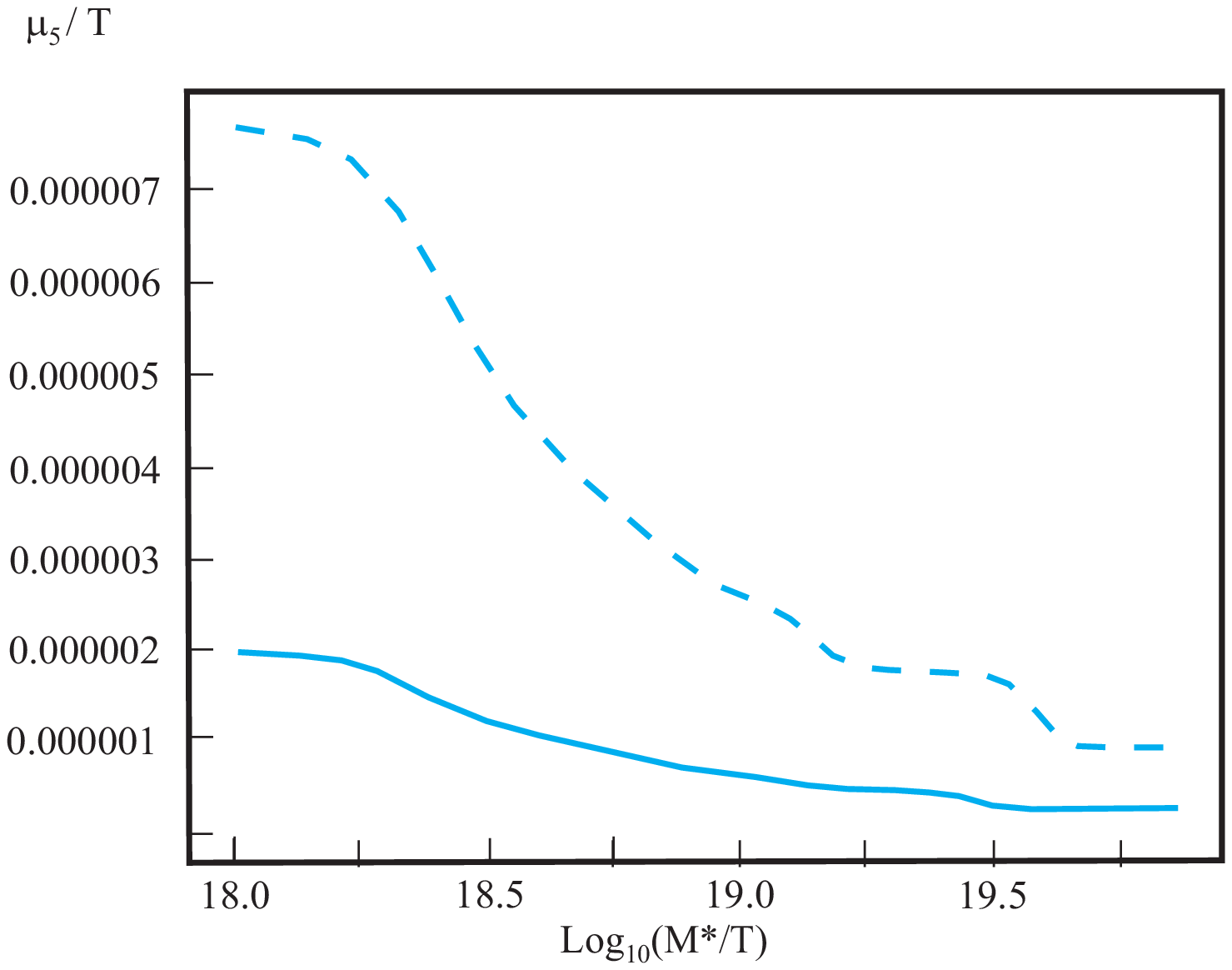}\\
\caption{\label{fig:H_homo_compar} Left panel: The Fourier components of the magnetic helicity in a homogeneous 
chiral plasma found in the model described by Eqs.~\eqref{B_k_disc_main}--\eqref{mu_5_k_disc_main} 
(solid lines) and in the model described by Eq.~\eqref{scalar-H}--\eqref{scalar-mu} (dashed lines). 
Right panel: The chiral chemical potential as a function of the (conformal) time in a homogeneous chiral plasma 
described by the same two models.}
\label{fig:Mu5_homo} 
\end{figure}

To test the model, we start by considering the case of a homogeneous chiral plasma. In this case, the only nonzero  
component of the chiral chemical potential is $\mu_{5,0}$. Our numerical results for the magnetic helicity modes and
the chiral chemical potential as functions of the (conformal) time are shown by the solid lines in Fig.~\ref{fig:H_homo_compar}. 
It is instructive to compare our results with those obtained from Eqs.~\eqref{scalar-H} and \eqref{scalar-mu}. Recall 
that the latter system of equations was used in the original analysis of the inverse cascade in Ref.~\cite{Boyarsky}. 
The corresponding results are shown by the dashed lines in Fig.~\ref{fig:H_homo_compar}. As is easy to see, the
results in both schemes have very similar qualitative behavior. Some quantitative differences should not be surprising
after one takes into account that very different sets of wave-vectors are used in the two analyses. Indeed, by following 
the approach of Ref.~\cite{Boyarsky}, a basis of spherical modes was used in solving Eqs.~(\ref{scalar-H}) and 
(\ref{scalar-mu}). In the numerical analysis of Eqs.~(\ref{B_k_disc_main}) and (\ref{mu_5_k_disc_main}), on the
other hand, a simpler basis of plane waves was utilized. 

\begin{figure}[t!]
\includegraphics[width=0.47\textwidth]{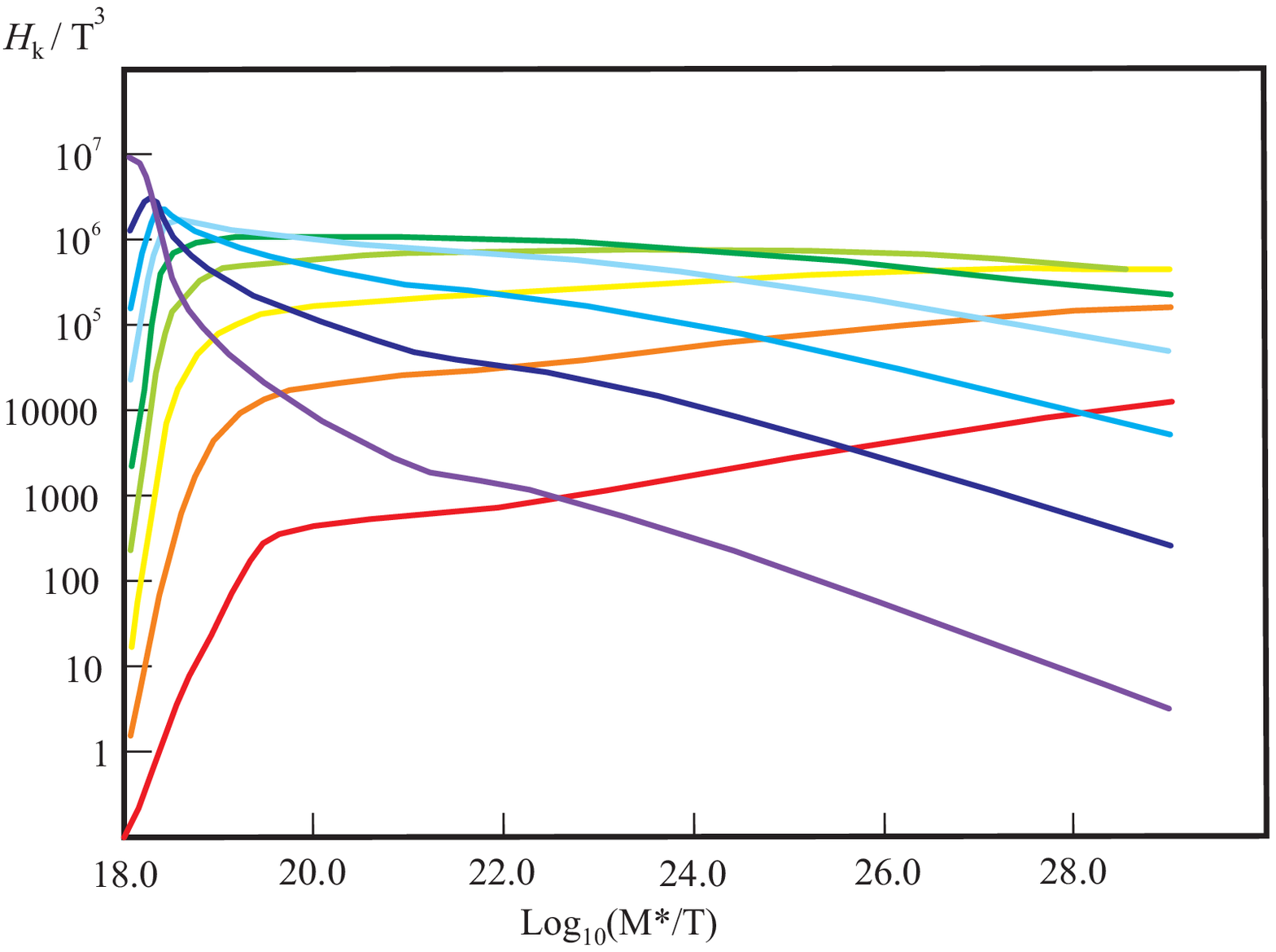}\hspace{0.05\textwidth}
\includegraphics[width=0.47\textwidth]{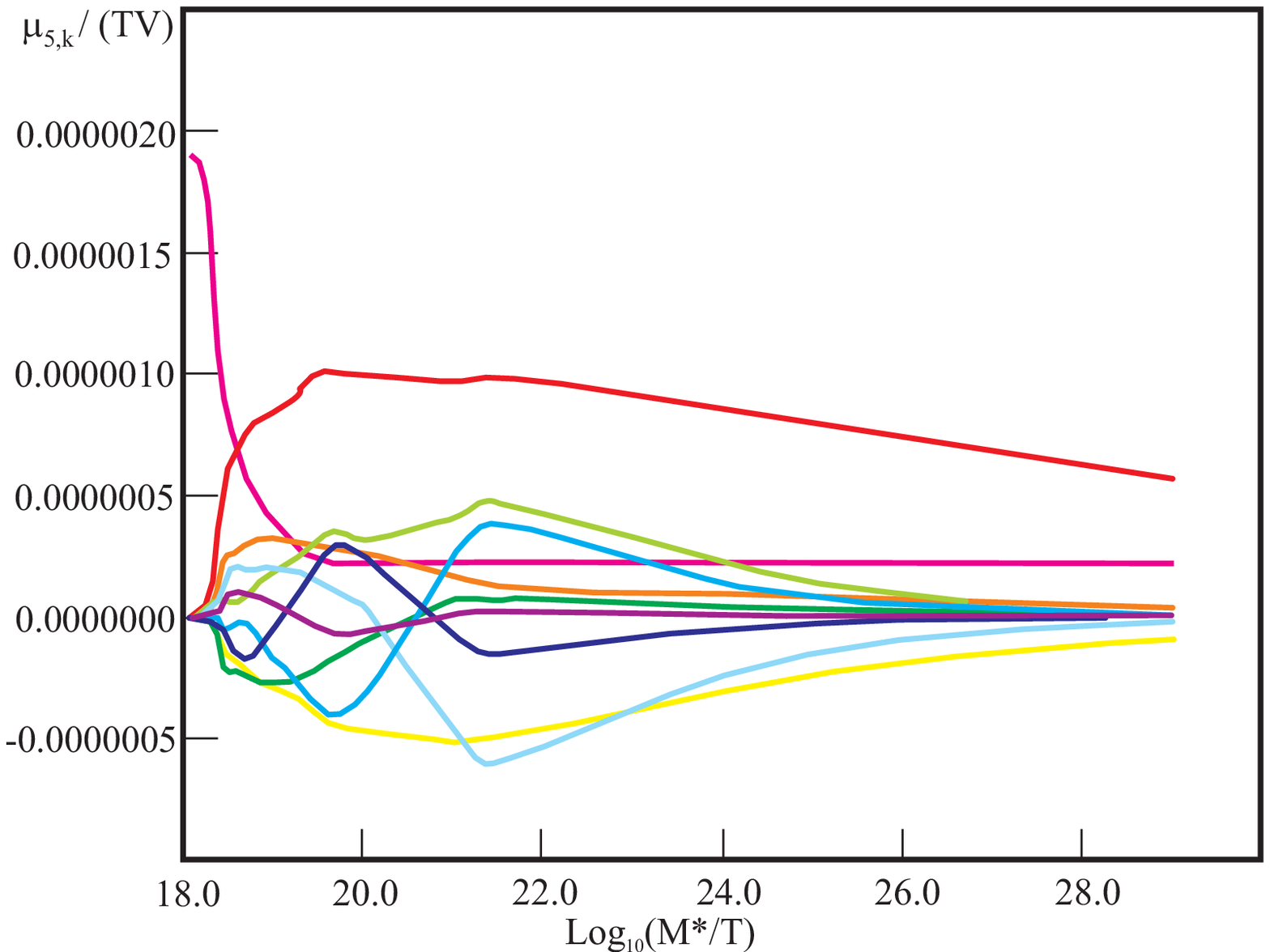}
\caption{The Fourier components of the magnetic helicity (left panel) and the Fourier components of the chiral chemical 
potential (right panel) as functions of the (conformal) time in an inhomogeneous chiral plasma. The results are obtained 
in a simplified model without diffusion, and the initial chiral asymmetry is assumed to be homogeneous.}
\label{fig:Inhomo_base}
\end{figure}

Now, let us proceed to the case of an inhomogeneous chiral plasma. Before considering the general case, 
we find it useful to study a toy model which ignores the effects of diffusion in the evolution of the chiral 
chemical potential. The corresponding effects are captured by the last term on the right hand side of 
Eq.~\eqref{mu_5_k_disc_main}. By taking into account that it is proportional to the square of the wave vector, 
diffusion will tend to diminish inhomogeneities. In order to address the role of $\mu_5$ inhomogeneities in 
the cleanest form possible, therefore, we start the analysis by ignoring the diffusion term. The corresponding 
results are presented in Fig.~\ref{fig:Inhomo_base}. The left panel shows the evolution of the magnetic helicity 
and the right panel shows the Fourier modes of the chiral chemical potential. (Note that, in view of the relation 
between the chiral chemical potential in coordinate and momentum space given in Eq.~(\ref{fourier_mu_5_x}), 
it more convenient to plot the momentum modes of the chiral chemical potential divided by $V$ rather 
than $\mu_{5,\textbf{k}}$.) For the initial conditions, we assumed that $\mu_{5,\textbf{0}}$ was the only nontrivial 
component, while all inhomogeneous modes $\mu_{5,k_z}$ (with $k_z=2\pi n_z/L$, where $n_z=1,...,N-1$) were set to 
zero at $t=0$. As is evident from the left panel of Fig.~\ref{fig:Inhomo_base}, the magnetic helicity is transferred from 
modes with shorter to longer wave lengths. This is nothing else, but the celebrated inverse cascade in the 
corresponding inhomogeneous plasma. An interesting feature of the inverse cascade here is that the 
inhomogeneous modes of the chiral chemical potential evolve in a rather nontrivial way: they grow first
in magnitude and then slowly decay later. 

\begin{figure}[t!]
\includegraphics[width=0.32\textwidth]{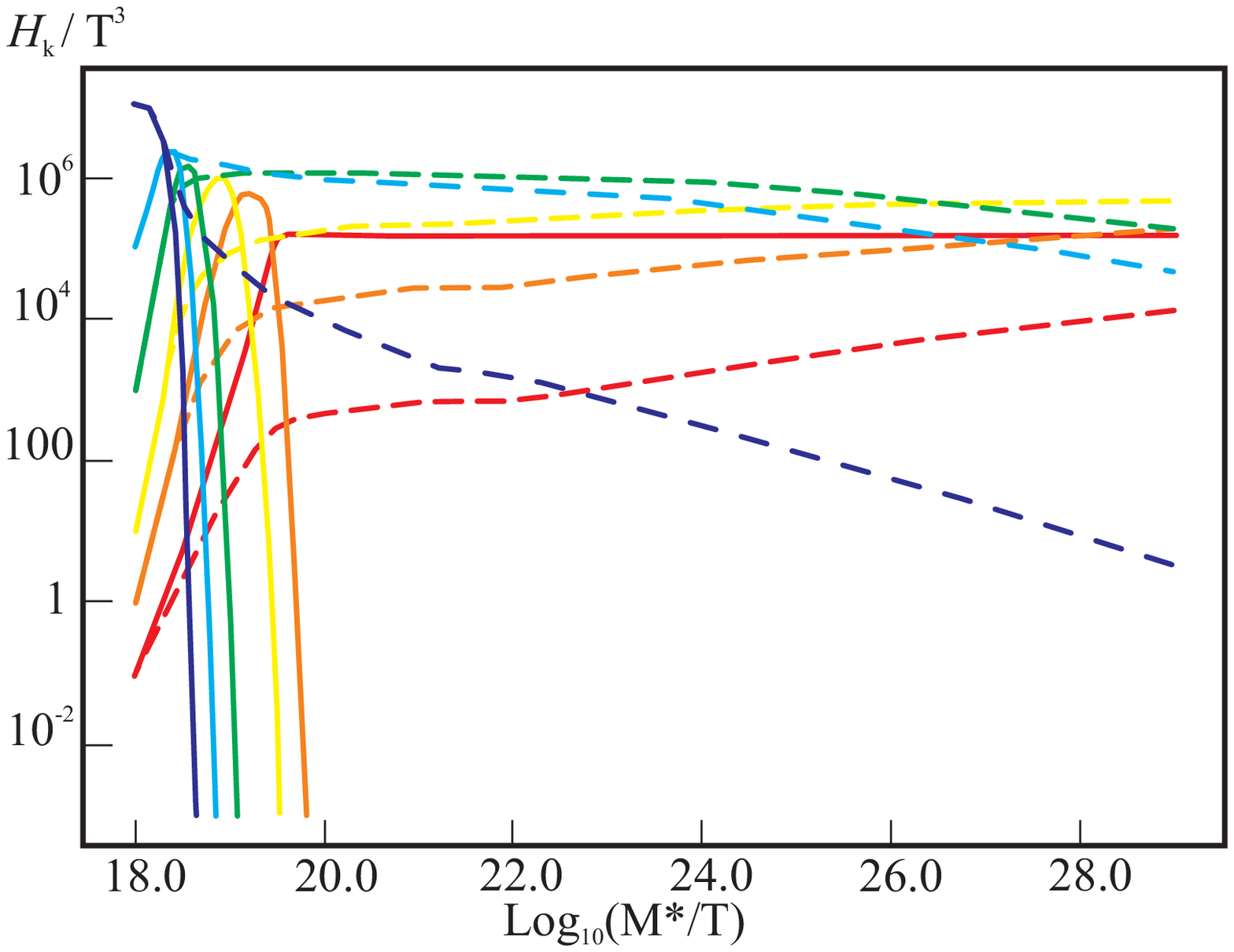}
\includegraphics[width=0.32\textwidth]{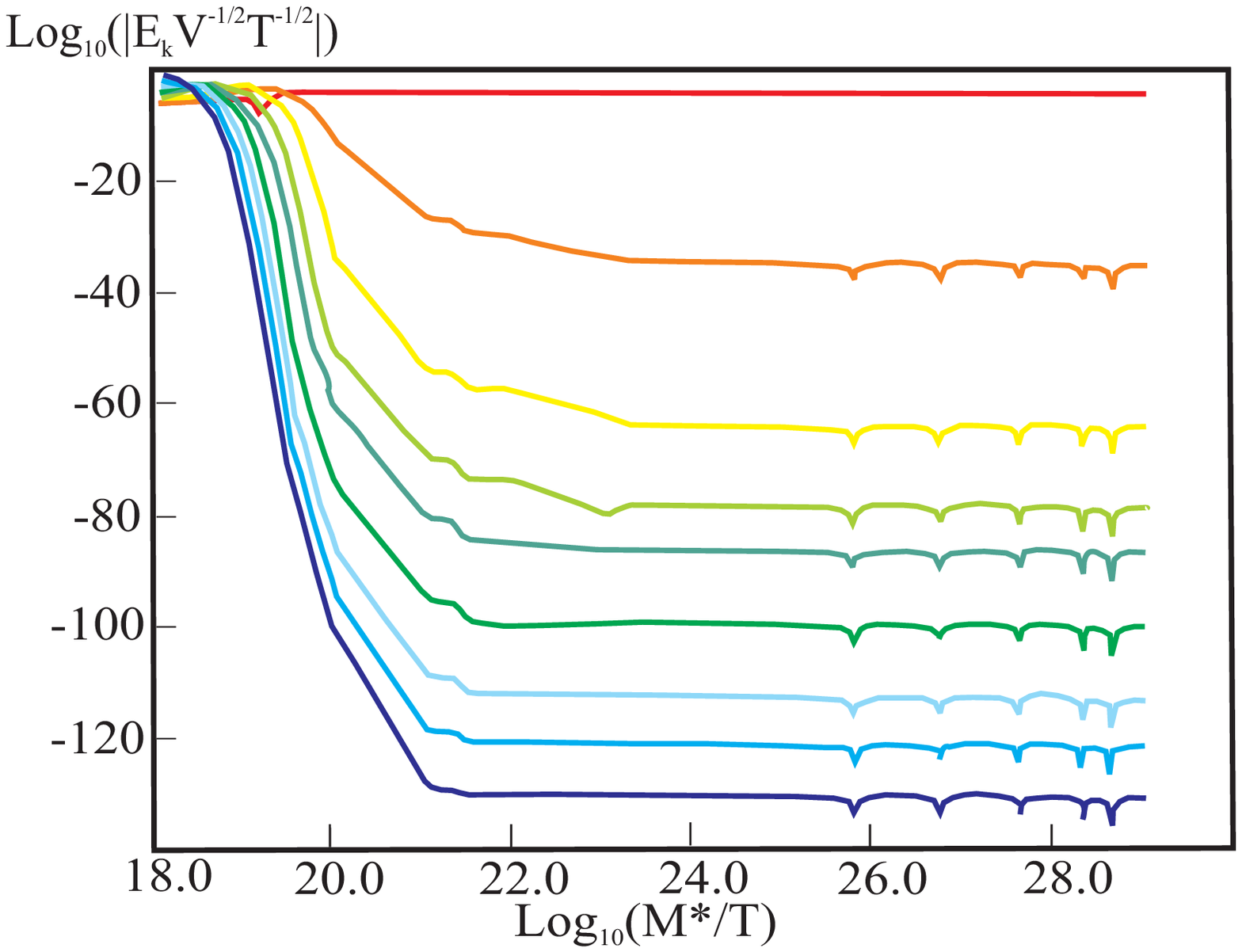}\vspace{0.05\textwidth}
\includegraphics[width=0.32\textwidth]{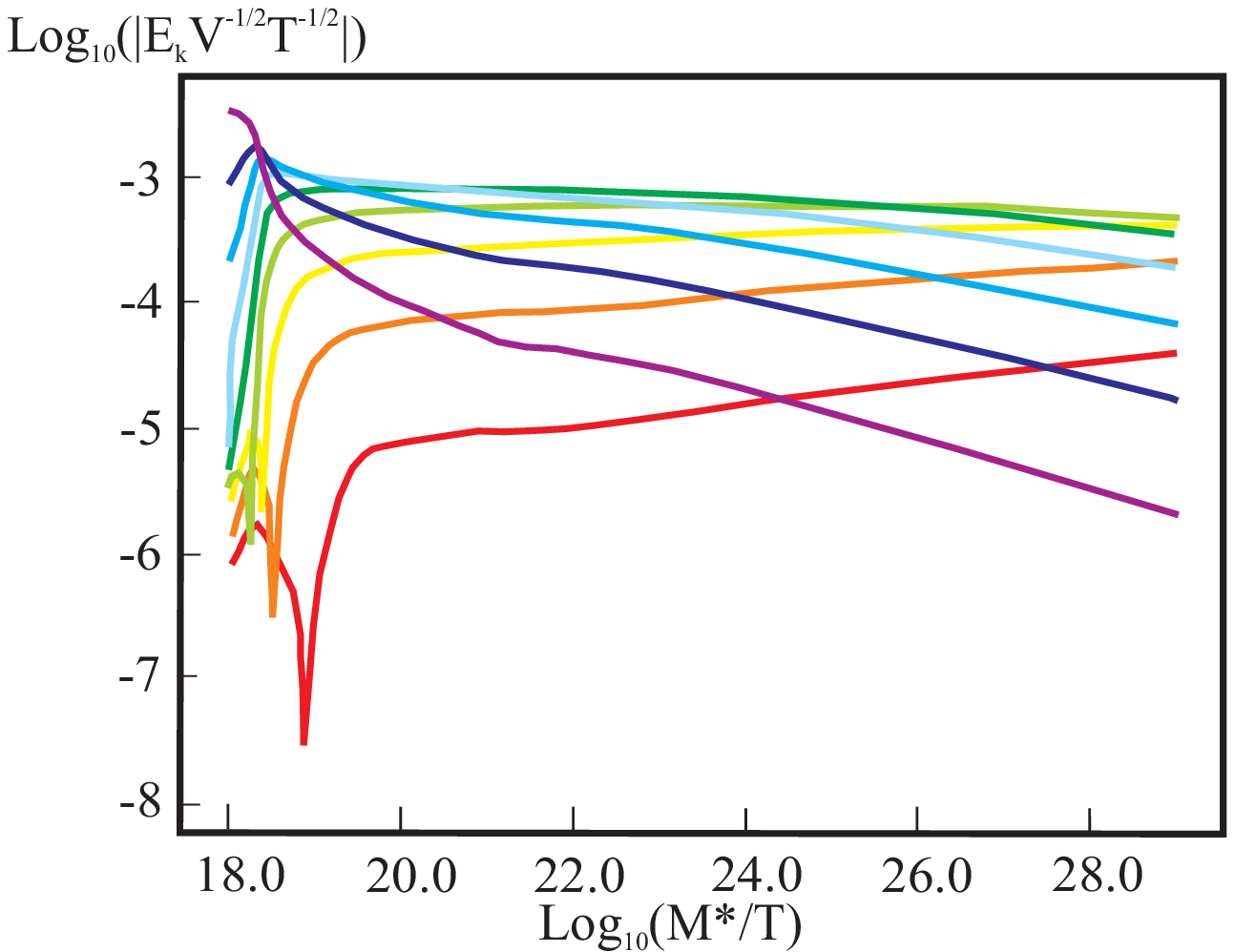}
\caption{The comparison of the magnetic helicity evolution in the homogeneous (solid lines) and inhomogeneous 
(dashed lines) models without diffusion (left panel), as well as the evolution of the electric field modes $|E_\textbf{k}|$ 
in the homogeneous (middle panel) and inhomogeneous (right panel) models without diffusion.}
\label{fig:H_compare} 
\end{figure}

We also observe that the rate of inverse cascade considerably slows down in the case with inhomogeneities 
of the chiral chemical potential included. This becomes obvious by comparing the evolution of the 
magnetic helicity modes in the homogeneous and inhomogeneous cases, shown in the left panel of Fig.~\ref{fig:H_compare} 
by the solid and dashed lines, respectively. As we see, the inverse cascade in the inhomogeneous case is 
much slower. The same is also evident from the evolution of the electric field, which is the driving force for the
inverse cascade. The corresponding results are also shown in Fig.~\ref{fig:H_compare} for homogeneous (middle panel) 
and inhomogeneous (right panel) chiral plasmas. The difference in the scales on the vertical axes should be 
noted. In agreement with the idea of a prolonged inverse cascade in the inhomogeneous case, the electric field 
modes remain substantial for a much longer time. It is important to emphasize, however, that the effects of diffusion 
were not included in the analysis yet. 

\begin{figure}[ht!]
\includegraphics[width=0.47\textwidth]{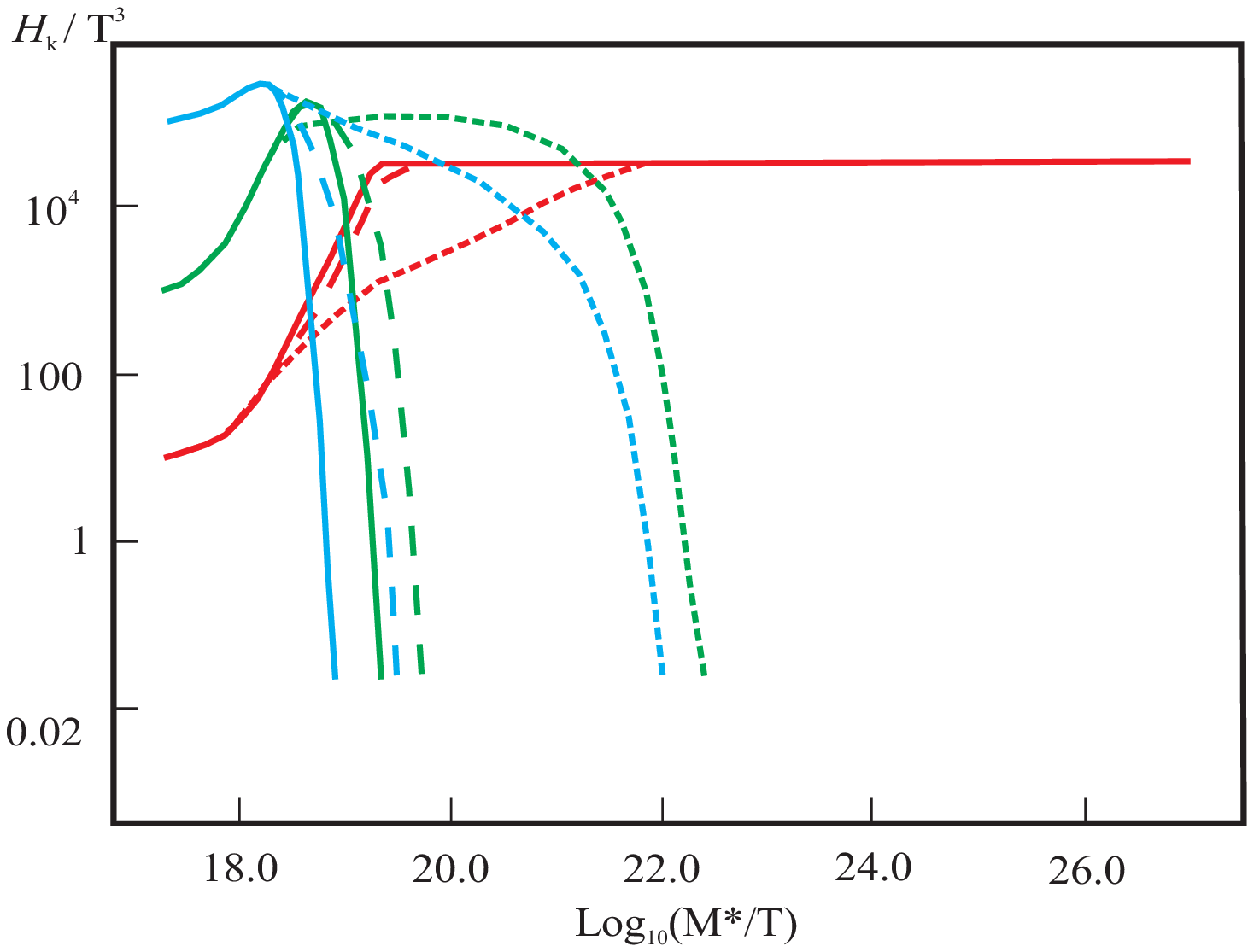}\hspace{0.05\textwidth}
\includegraphics[width=0.47\textwidth]{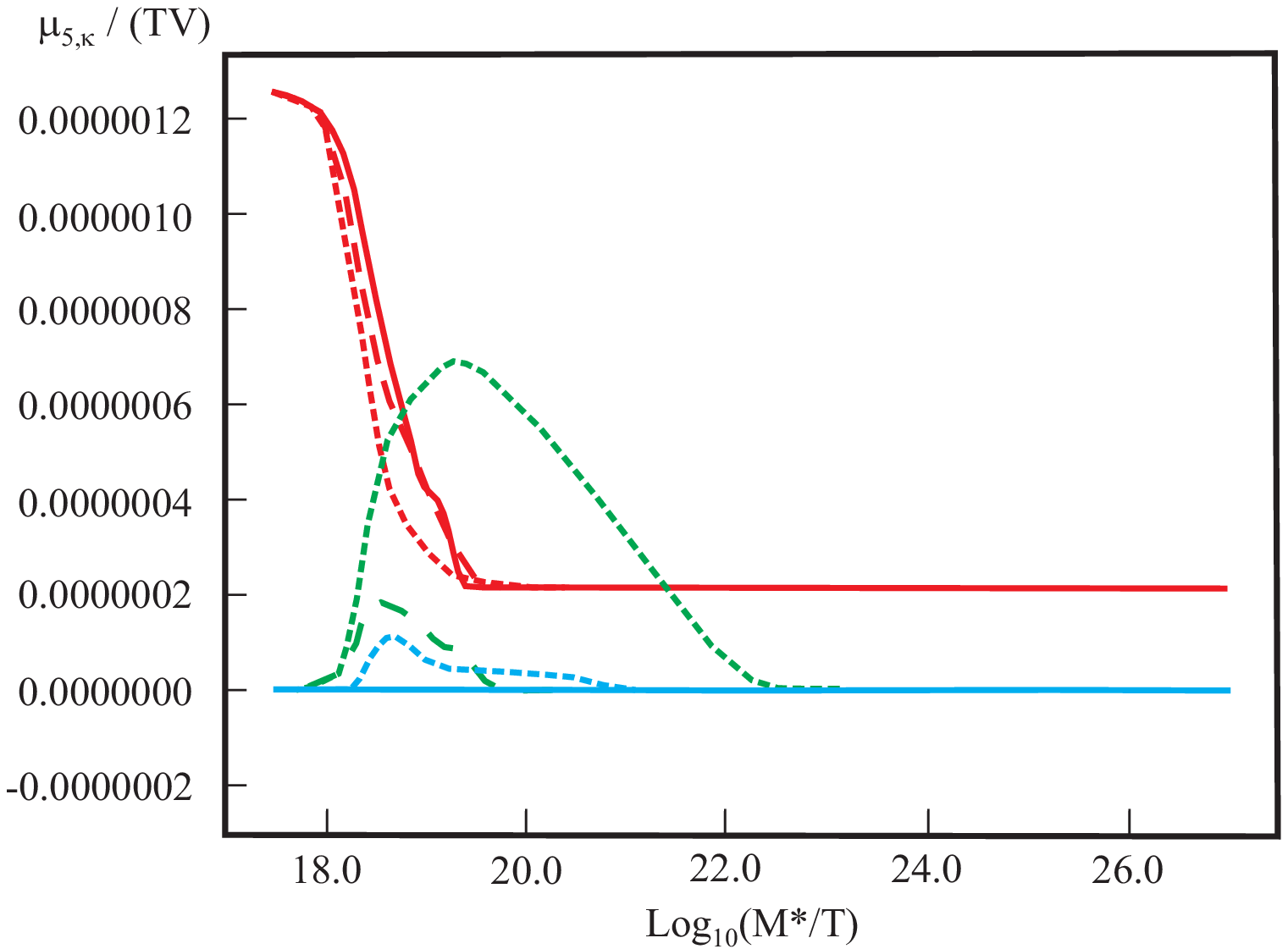}
\caption{The Fourier components of the magnetic helicity (left panel) and the Fourier components of the chiral chemical 
potential (right panel) as functions of the (conformal) time in an inhomogeneous chiral plasma with the effects of diffusion included. 
The results are shown for three values of the dimensionless relaxation time $\tau=6.8\times 10^3$ (solid lines), $\tau=6.8$ (dashed lines), and 
$\tau=6.8\times 10^{-3}$ (dotted lines). For clarity of the presentation, only three modes are included.}
\label{fig:evolution-diffusion}
\end{figure}

With a clear understanding of the role that inhomogeneities of the chiral chemical potential play in the dynamics 
without diffusion, it is now important to study how the diffusion term affects the inverse cascade. Our numerical 
results for the magnetic helicity in the diffusive dynamics are presented in the left panel of Fig.~\ref{fig:evolution-diffusion}.
In order to get a deeper insight into the underlying physics, we consider several values of the 
dimensionless relaxation time ($\tau\to \tau T$) that control the magnitude of the diffusion term: 
$\tau= 6.8\times 10^3 $, which corresponds to the dimensionless conductivity of the primordial plasma
$\sigma \approx 70$ \cite{Baym}, as well as two smaller relaxation times, $\tau=6.8$ and $\tau=6.8\times 10^{-3}$. 
For a sufficiently small relaxation time $\tau$, the inverse cascade slows down just like in the absence of diffusion. 
For the realistic value of the relaxation time in the primordial plasma, however,  the evolution of the magnetic helicity 
does not differ much from the homogeneous case. This is due to the fact that the modes of the chiral chemical 
potential with nonzero wave vectors appear to be rather strongly suppressed and play little role in the evolution. 

It is interesting to mention that relativistic chiral plasma can be also realized in table-top experiments. 
Indeed, charge carriers in Dirac/Weyl materials are described by massless spin-$1/2$ particles \cite{Vafek:2013mpa}. 
Thus, the corresponding quasiparticle plasma could mimic the hot plasma in the early Universe, as well 
as cold dense matter of hypothetical quark stars. It would be interesting, therefore, to explore experimentally if 
a version of the magnetic helicity evolution studied in this paper could be realized in the Dirac/Weyl 
(semi-)metals. Naively, this should be possible indeed \cite{Hirono:2015rla}.

\begin{figure}[ht!]
\includegraphics[width=0.46\textwidth]{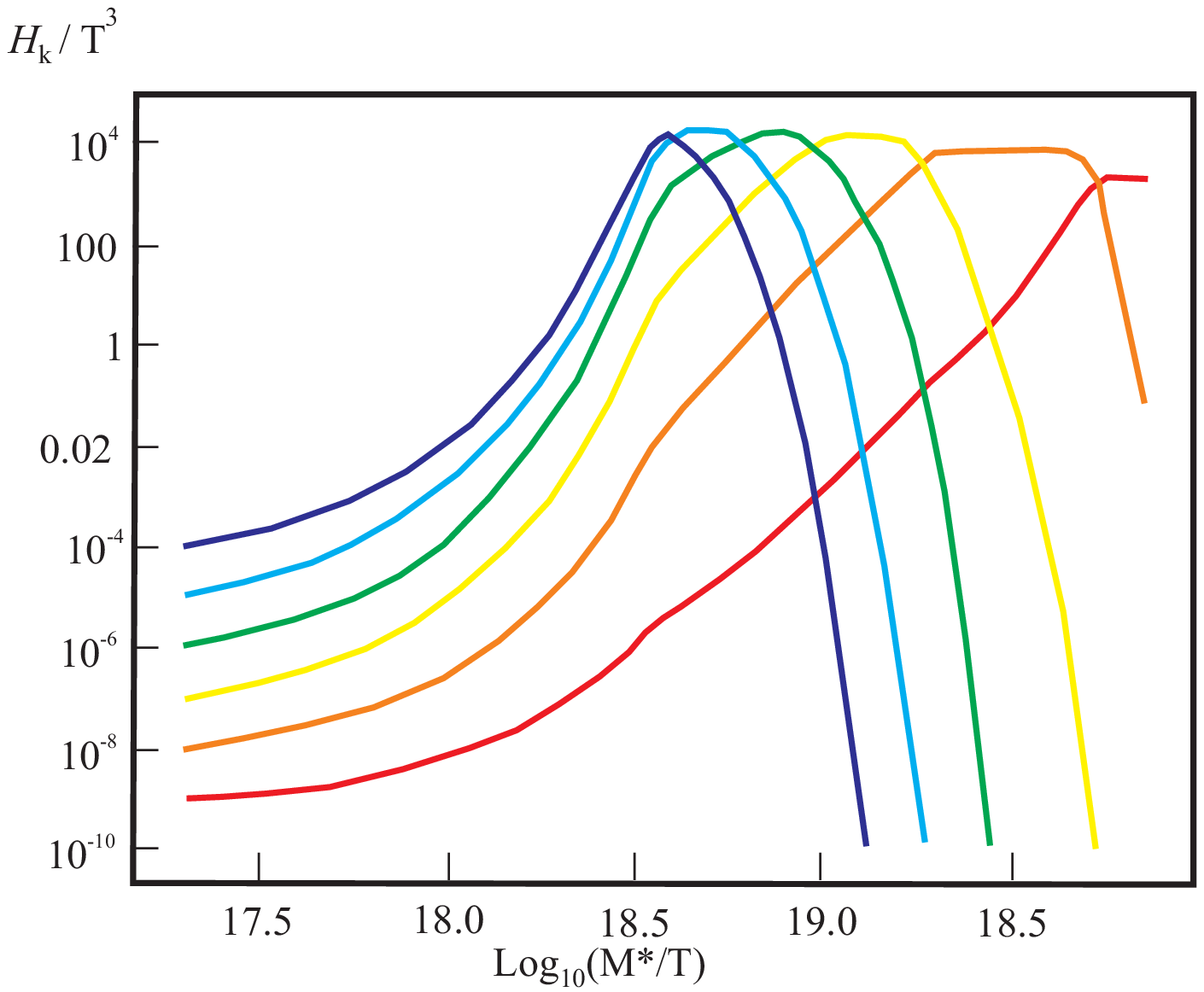}\hspace{0.05\textwidth}
\includegraphics[width=0.48\textwidth]{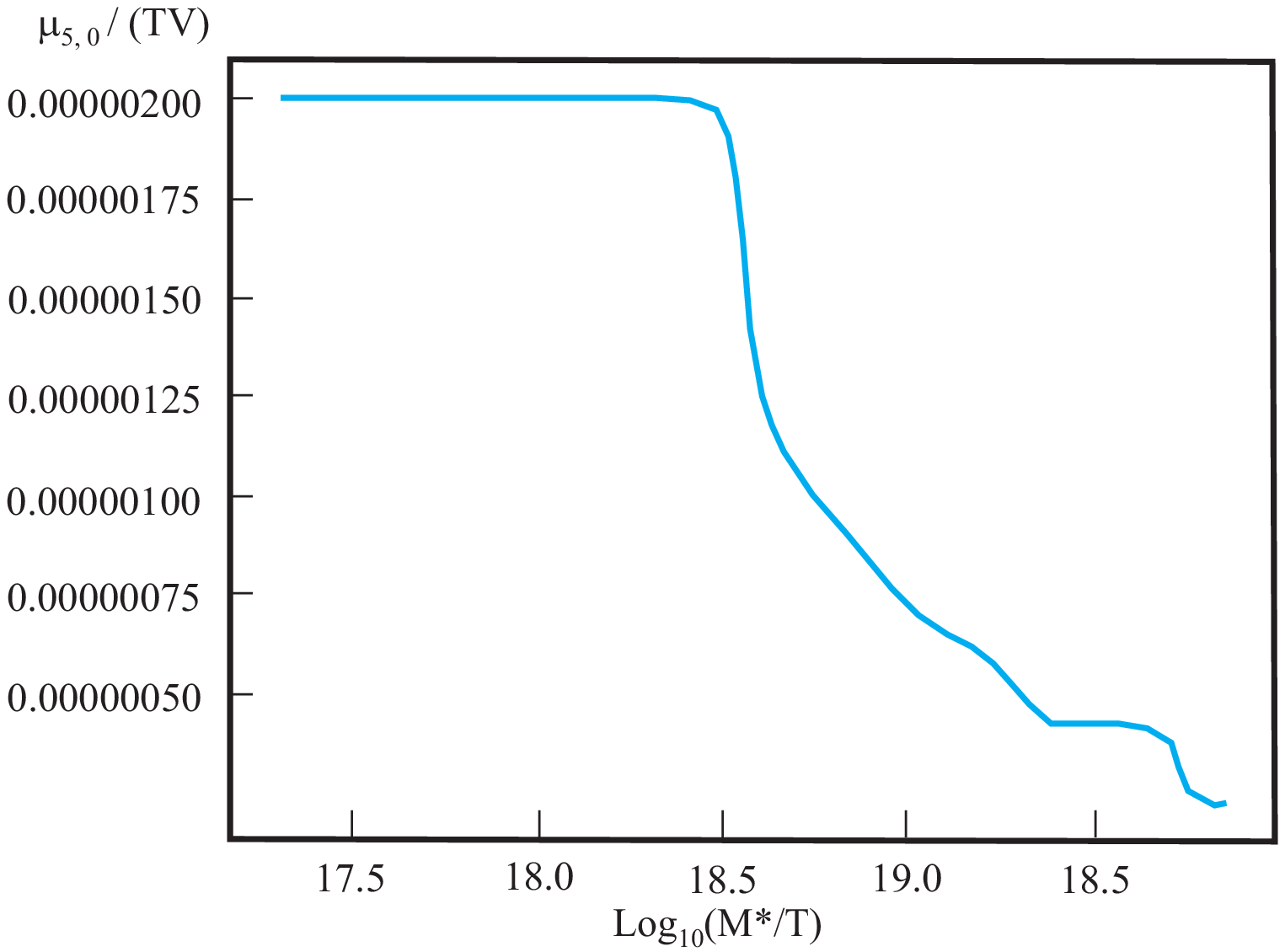}
\caption{The Fourier components of the magnetic helicity (left panel) and the chiral chemical potential (right panel) 
as functions of the (conformal) time in the case of very small initial magnetic fields, with the effects of diffusion 
included ($\tau=6.8\times 10^3$).}
\label{fig:evolution-small_H}
\end{figure}

A nonzero chiral asymmetry in a Dirac/Weyl material could be induced via the chiral separation effect \cite{Vilenkin:80a,Metlitski:2005pr}. 
Perhaps, a nonzero magnetic helicity can be also induced by applying certain configurations of external electric and magnetic 
fields, or special types of electromagnetic waves. The latter, however, may be difficult in a metallic regime. 
Nevertheless, we argue that the inverse cascade can be easily triggered by creating an initial nonzero chiral 
asymmetry. This is supported by the numerical study of Eqs.~\eqref{B_k_disc_main}--\eqref{mu_5_k_disc_main} 
in the case with very small initial magnetic fields, but a nonzero chiral chemical potential. The corresponding 
results are shown in Fig.~\ref{fig:evolution-small_H}. Note that, because of large diffusion ($\tau=6.8\times 10^3$), 
the Fourier components of the chiral chemical potential with $\mathbf{k}\neq 0$ remain negligible during the whole 
period of the evolution.

\section{Conclusion}
\label{Conclusion}

The essence of the inverse cascade is that the magnetic helicity is transferred rather 
efficiently from helical magnetic modes with short wavelengths to long ones \cite{Boyarsky,Tashiro:2012mf,Manuel:2015zpa,Hirono:2015rla}. 
It should be noted that the key role in the underlying dynamics is played by the anomalous effects associated with the chiral 
symmetry. The previous studies used model frameworks in which the feedback of the anomalous dynamics was treated via
a space-average chiral anomaly relation. By construction, such a treatment did not allow for inhomogeneities in the chiral 
asymmetry to develop. This is a limitation, of course, because some inhomogeneities would seem unavoidable (e.g., as 
pre-existing and/or dynamically developed fluctuations). Moreover, one could naively speculate that the corresponding 
inhomogeneities may even quench the inverse cascade from developing \cite{inhomogeneous}.

In this paper, we studied how the diffusion and inhomogeneities associated with the chiral asymmetry affect the inverse 
cascade scenario in a chiral relativistic plasma. We used a simple toy model that captures the anomalous dynamics and 
showed that the inverse cascade still survives in a plasma with dynamical inhomogeneities of the chiral asymmetry. We 
find that the outcome of the underlying dynamics is sensitive to the effects of diffusion. When diffusion is negligible, the 
inverse cascade evolves much slower than in the model with a homogeneous chiral charge density. When the diffusion is large, 
on the other hand, the inhomogeneities are very efficiently suppressed and the inverse cascade becomes indistinguishable 
from that in the chirally homogeneous model. By taking into account the realistic value of the diffusion term in the primordial plasma, 
we conclude that the corresponding dynamics is highly diffusing in the early Universe and, thus, the inhomogeneities of
the chiral asymmetry play little role. 

Finally, it would be interesting to study the anomaly-driven inverse cascade in Dirac and Weyl metals \cite{Vafek:2013mpa}, whose low 
energy excitations are described by the Dirac and Weyl equations, respectively. As shown by Nielsen and Ninomiya \cite{Nielsen-Ninomiya},
one may generate a chiral asymmetry in such condensed matter systems by applying the electric and magnetic fields in the same direction.
If the magnetic field is spatially non-uniform, an inhomogeneous chiral asymmetry will be generated. Switching off the external electric and
magnetic fields, one then can study how this chiral asymmetry and the helical magnetic fields generated by this asymmetry will evolve in time
and whether an inverse cascade will result.

\acknowledgments
The authors would like to thank A.~Boyarsky and O.~Ruchayskiy for fruitful discussions. 
This work was supported in part by the Swiss National Science Foundation, Grant No. SCOPE IZ7370-152581.
E.V.G. is grateful to the Program of Fundamental Research of the Physics and Astronomy 
Division of the NAS of Ukraine for the support. The work of I.A.S. was supported by the U.S. 
National Science Foundation under Grant No.~PHY-1404232. S.V. is grateful to the Swiss 
National Science Foundation (individual Grant No.~IZKOZ2-154984).

\end{document}